# Modelling effective electrical resistance in particle reinforced composites using Generative Adversarial Network

Vinit V Deshpande, Pascal A Happ, Romana Piat

*Department of Mathematics and Natural Sciences, University of Applied Sciences, Darmstadt, Germany*

**Abstract:** Polymer matrix composites embedded with conductive particles are widely utilized for applications that demand stringent control of the effective electrical resistance (or conductivity) of the material. This property is highly sensitive to the particle shape and size distribution within the composite and their percolation threshold. One of the most widely utilized numerical strategies to model this property is the Resistor Network method. However, it is based on many assumptions of the particle shape and inter-particle contact which limits its practical applications. In this work, we have proposed a conditional Generative Adversarial Network (cGAN) based modelling strategy that can accurately capture the flow of electrical current through particles which are connected to multiple other particles in the matrix. The cGAN is trained on data generated by finite element simulations that can model the physics of the problem accurately. It is shown that the GAN based model predicts the electrical flow within the composite and hence the effective electrical resistance much more accurately than the Resistor Network model.

## 1. Introduction

Conducting polymer composites incorporate electrically conducting particles in polymer matrix with an objective of making the otherwise insulating material conductive while taking advantage of moulding ability of the polymer. The generated composite material can be moulded into different shapes and sizes and can have tailored electrical conductivity [1]. The effective electrical conductivity of such composites depends on the conductivity of the conducting material, shape of the conductive particles, their volume fraction, their dispersion in the polymer matrix and the percolation threshold. Each of these parameters can be controlled to obtain application specific conductivity. Typical conducting materials used are carbon black (CB) [2], graphene [3], carbon nanotubes (CNTs) [4], metal nanoparticles [5], etc.

Such composites are used in a wide variety of applications like strain sensing [6], human-machine interaction [7], soft robotics [8], structural health monitoring [9], human motion detection [10], etc. These applications involve detecting mechanical response in terms of strain from the change in resistance (or conductivity) of the material. Each of these applications have different requirements for measurement sensitivity also called as gauge factor (GF). [11] developed a CB-CNTs based polymer composite for human motion monitoring with a GF between 0.91 to 13.1. Porous graphene nanoplates based polymer composites were developed in [12] with a GF between 1.9 to 17.8. [13] produced CB reinforced composites utilizing different polymers like thermoplastic polyurethane and olefin block copolymer to develop stretchable sensors with GFs in range of 1 to 50 for the strain range lower than 50%. The consideration of GF in the present work of numerical modelling is important because the prediction of any numerical model should have accuracy much higher than the GF. In other words, the relative error between the prediction of electrical conductivity from any numerical model and that of a reference method (either experimental measurement or a well-accepted numerical method) should be much smaller than the GF of the application.

Two types of numerical strategies are well-accepted to model the effective electrical conductivity in particle-reinforced composites. One is approximate solutions to the Laplace equation, $\nabla^2 \phi = 0$ where $\phi$ is the potential function. [14] developed finite element method (FEM) based procedure that modelled the particle geometry and inter-particle interactions accurately. [15] utilized finite volume method (FVM) in combination with discrete element method (DEM) to create the microstructure and solve the problem. These methods are accurate in modelling the intra-particle resistance because of their ability to capture the particle geometry perfectly. However, the computational expense of these methods is very high which makes it difficult to use them to model materials at the scale of real applications.

In order to deal with these limitations, an approximate method called 'Resistor Network' is utilized which models the particles as one-dimensional resistors which reduces the computational expense significantly. [15] utilized resistor networks to calculate effective resistance of a CB based composite. [16] modelled microstructure of an all-solid-state battery using resistor networks. A big challenge in utilizing these methods is how to model the intra-particle resistance. [16] assumed the shape of the particle as a cylinder to determine is effective resistance which drastically overestimated the intra-particle resistance. [15] utilized an analytical solution to the Laplace equation for a spherical particle given certain specific boundary condition. This solution works well in case a particle is in contact with only two other particles but it fails drastically if there are more than two neighbouring particles. In real microstructures, the particles that form a conducting path have multiple neighbouring particles which makes using the above method unsuitable. Modelling intra-particle resistance is an unsolved problem especially for particles that have complex shapes and arbitrary contact areas with other particles as is seen in most composites. The assumption of cylindrical shape of particles or analytical solution of Laplace equation leads to significant errors in the calculation of the effective resistivity of the individual particles and hence that of the entire composite.

In this work, a generative AI based method is developed to model current flow through a 2D microstructure. Generative Adversarial Network (GAN) is used to predict current flow through a given microstructure. Its prediction is compared against that of a FEM prediction and the best available Resistor Network model. It will be showed that prediction of the GAN is as good as that of the FEM simulation and much better than the Resistor Network model.

## 2. Single particle studies

### 2.1 Analytical solution

Calculation of effective electrical resistance between two contact points of a disk was first discussed in [16] and its derivation can be found in [17]. Consider a disk shown in Fig.1. Let d be the distance between two electrical contacts represented by cylindrical electrodes of diameter $\delta$.

$\phi(p)$ is electrical potential at any point p in the material. It is located at a distance $r_1$ from electrical contact 1 having charge density $\lambda$ and $r_2$ from electrical contact 2 having charge density $-\lambda$. The electrical potential at point p is given as,

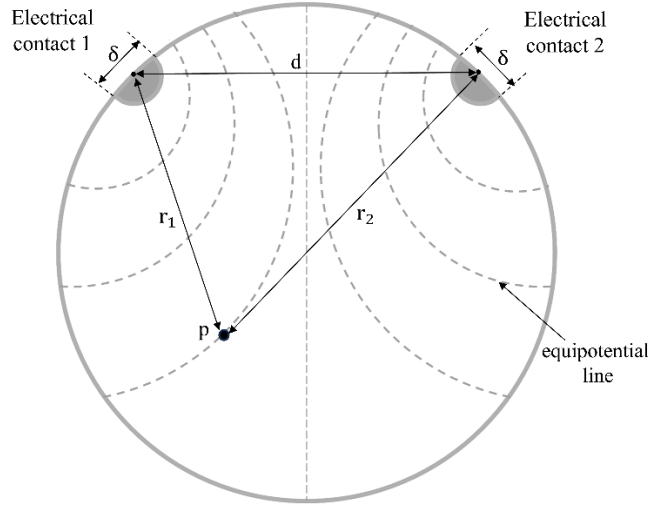

Fig.1 A 2D circular disk connected to two electrodes with circular contact areas [17].

$$\phi(p) = 2\lambda \ln \frac{r_1}{r_2} \qquad (1)$$

where $r_1, r_2 \geq \frac{\delta}{2}$. The above equation shows that the equipotential lines are circular. The potential at any point p on the surface of electrode 1 is,

$$\phi(\text{surface 1}) = 2\lambda \ln \frac{d - \delta/2}{\delta/2} \approx 2\lambda \ln \frac{2d}{\delta} \qquad (2)$$

The potential at any point p on the surface of electrode 2 is negative of that of electrode 1. Therefore, the potential difference between the two electrodes is,

$$V \approx 4\lambda \ln \frac{2d}{\delta} \qquad (3)$$

The current I can be calculated by integrating current density J along am equipotential surface. Consider an equipotential surface located very close to the contact surface 1. So, $r \ll d$ and the electric field E on this surface has magnitude,

$$E = \frac{2\lambda}{r} \qquad (4)$$

The current density across this surface is given by $J = \sigma E$ where $\sigma$ is the conductivity of the material. Considering t as thickness of the disk, the total current flowing across this surface is,

$$I = J.A = \frac{2\sigma\lambda}{r} . \pi rt = 2\pi\sigma\lambda t . \tag{5}$$

Finally, the resistance across the two contact points is,

$$R = \frac{V}{I} \approx \frac{4\lambda \ln\frac{2d}{\delta}}{2\pi\sigma\lambda t} = \frac{2}{\pi\sigma t} \ln\frac{2d}{\delta} . \tag{6}$$

Note that, the both the contact areas are circular and the equipotential lines are also circular. It will be shown next that the results deviate significantly if the contact areas re non-circular as will be the case in most practical applications.

## 2.2 Finite element simulation

A finite element simulation was performed using Abaqus software. It solved the Laplace equation $\nabla^2 \phi = 0$ for a carbon black disk of radius 0.05mm, thickness 0.01mm, conductivity $\sigma = 0.1$ S/mm. Circular contact areas of diameter $\delta = 0.001$mm were created where are voltage of 1V is applied to one surface and 0V to another surface. The geometry is meshed using a 4-node linear quadrilateral element.

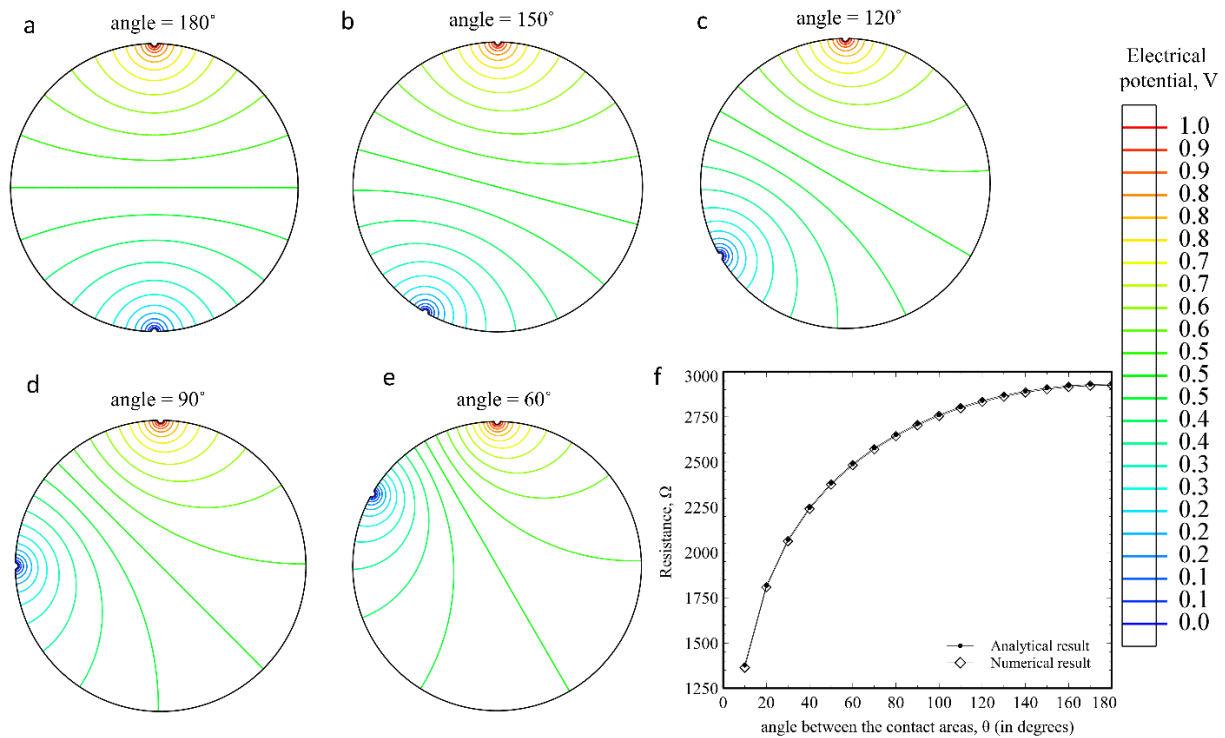

Fig.2. The equipotential lines in the disk for angles of a) 180°, b) 150°, c) 120°, d) 90° and e) 60° between the two contact areas; f) resistance vs. angle obtained from analytical calculations of section 2.1 and FEM simulations.

Fig.2 shows the equipotential lines calculated from FEM simulation for different angles between the contact areas. The angle between the two contact areas $\theta$ is related to the distance d and disk radius $a$ as $d = 2a \sin \theta/2$. Fig.2f shows the relation between resistance and angle between the contact areas calculated from analytical expression of Eq.6 and FEM simulations. The relative error of analytical results w.r.t the FEM results are very low with 1 % for $\theta = 10°$ to 0.2 % for $\theta = 10°$.

Similar study was performed by changing the contact areas to flat surfaces or a line contacts in 2D of width 0.002mm. Fig.3 shows the equipotential lines for this case. The grey color lines correspond to circular contact areas as shown in Fig.2.

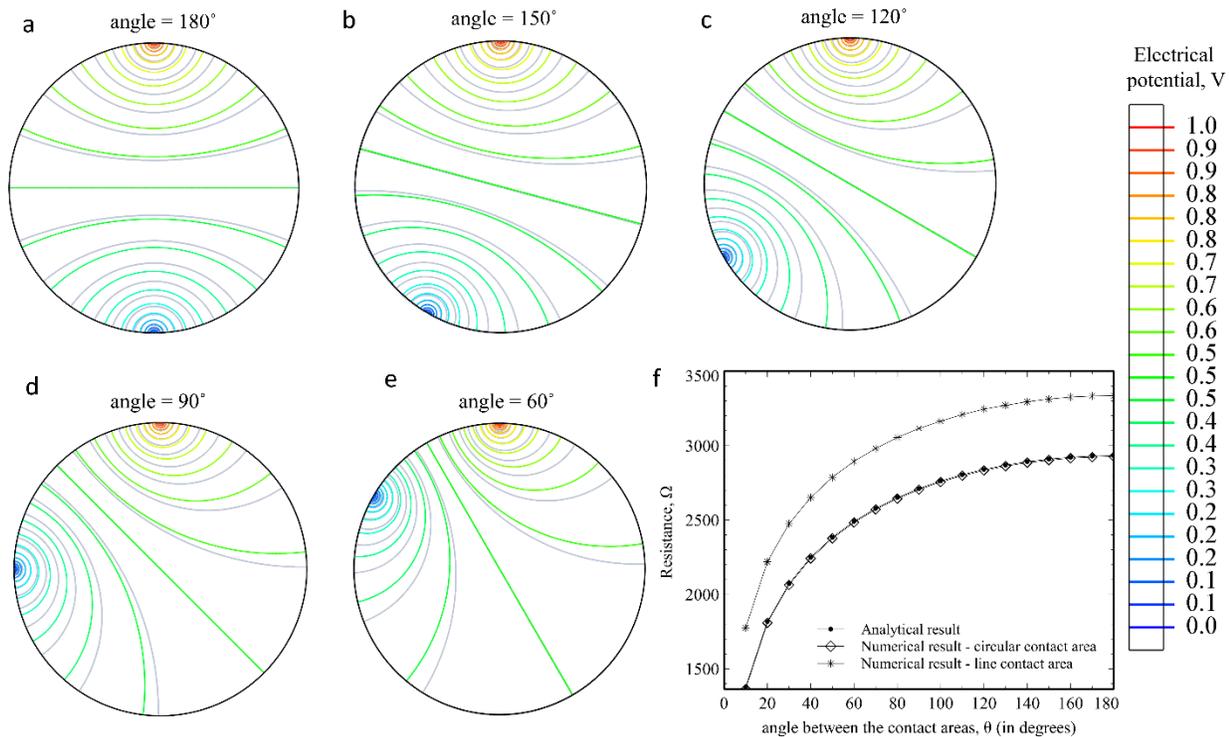

Fig.3. The equipotential lines in the disk for angles of a) 180˚, b) 150˚, c) 120˚, d) 90˚ and e) 60˚ between the two contact areas (grey color lines correspond to the case shown in Fig.2); f) resistance vs. angle obtained from analytical calculations of section 2.1 and FEM simulations for circular contact area and line contact area.

Fig.3f shows the relation between resistance and angle between the contact areas calculated from analytical expression of Eq.6 and FEM simulations of line contact and circular contact areas. The relative error of analytical results w.r.t the FEM results for line contact are very high with 22.4 % for $\theta = 10°$ to 12.1 % for $\theta = 180°$. The analytical results severely underestimate the resistance values. The relative error is in the same range as the values of GF of many composites as explained in the introduction. This means that the any prediction of effective resistance using the analytical expression is bound to mislead.

This shows that the nature of contact between the two particles drastically changes the effective resistance. Another point of consideration is that in real microstructures, the particles are often connected to more than two particles. The analytical expression works only in the case of two contacts on a single particle. Fig. 4 shows a particle that has three contact areas. 1V is applied to the top contact area and 0V to both the bottom contact areas. The angle between the top and the bottom contact areas is 140°.

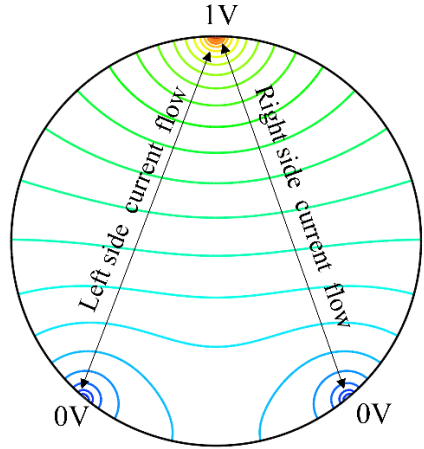

Fig.4. The equipotential lines in the disk with three contact areas; f)

Table 2. Resistance in the left side and the right side current flow (refer Fig.4) from FEM and analytical calculation along with the relative error between them.

| Resistance position | FEM results (Ω) | Analytical results(Ω) | % relative error |
| --- | --- | --- | --- |
| Left side | 5065.34 | 2892 | 42.9 |
| Right side | 5065.34 | 2892 | 42.9 |

It can be seen in Fig.4 that the equipotential lines take complex shapes in case of more than two contact areas. If the analytical expression is used to calculate the resistance of both left and right-side contact pairs in an uncoupled fashion then, it significantly underestimates the resistance with 42.9 % relative error. [15] utilized a similar analytical solution to model the resistances of multiple contacts of a spherical particle in an uncoupled fashion. But this study shows that this strategy leads to significant errors.

## 3. Multi particle studies

Some micrographs of CB reinforced polymer composites published in [18] can be seen in Fig.5. Since the matrix is an insulator, the particles that form a conducting path are the only ones that contribute to the effective conductivity of the composite. In this work, a 2D composite with circular reinforcing particles is utilized as a case study to compare the different numerical strategies.

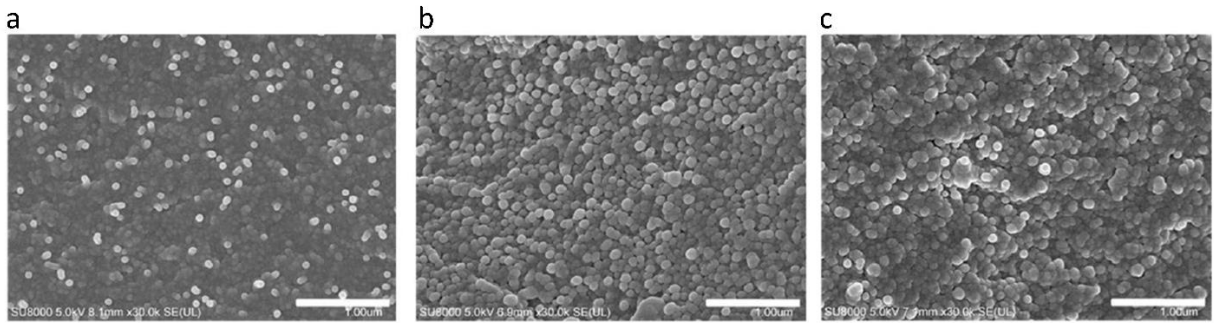

Fig.5. SEM micrograph of CB/PET with polymer content a) 0.1 wt % , b) 3 wt % and c) 10 wt % as published in [18].

### 3.1 Finite element simulations

Consider a square composite specimen with 1mm edge length and 0.01mmm thickness. The circular particles intersect each other with the depth of intersection chosen from a random distribution. These particles represent a conducting path. In real microstructures, there are many isolated particles that do not carry current and hence they are not required to model. The particles have conductivity of 0.1 S/mm and the matrix have conductivity of 8.5E-16 S/mm. 10V is applied to the top edge and 0V to the bottom edge.

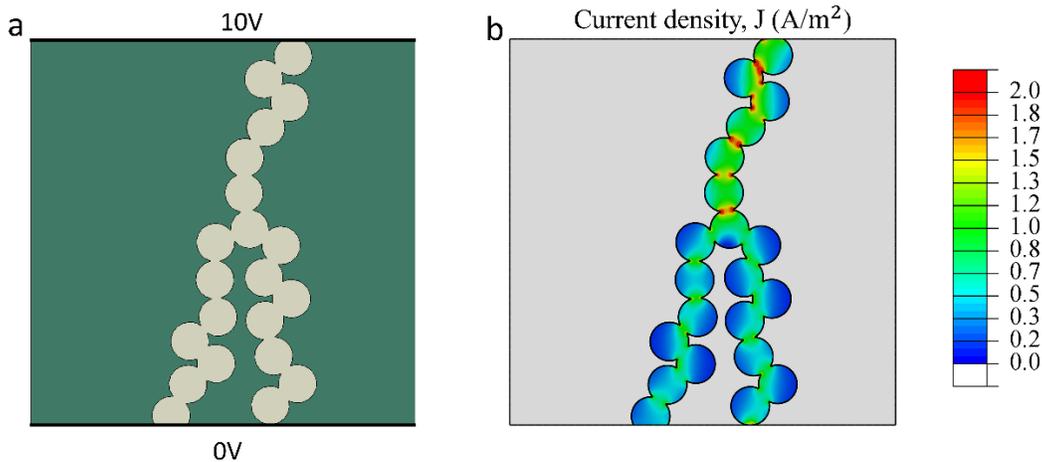

Fig.6. a) A 2D composite specimen with boundary conditions; b) contour of magnitude of current density in the particles.

The total current flowing through the composite is measured at the top edge (or at the bottom edge) and its value is, I = 7.6000e-04 A and the total resistance is, $R^{total}$ = $(\phi^{top\ edge} - \phi^{bottom\ edge})/I$ = 13158 Ω.

## 3.2 Resistor Network

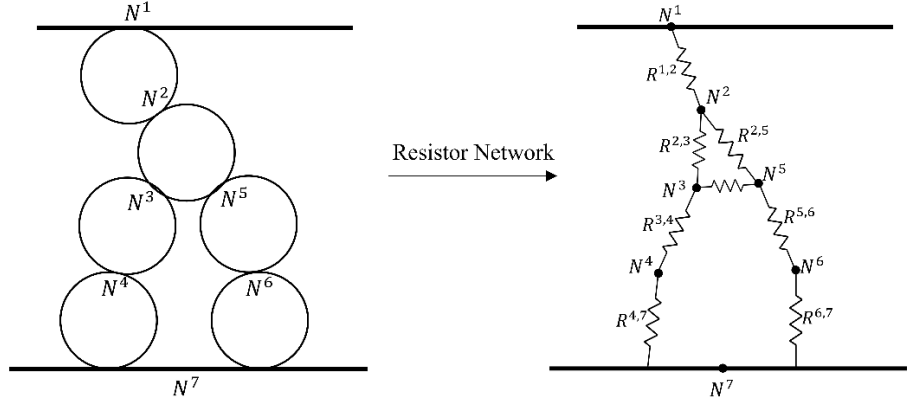

Fig.7. An illustration of a 2D composite specimen and its Resistor Network model.

Fig.7 illustrates how a Resistor Network is created. In a Resistor Network model, each contact $i$ between the particles is modelled as a node, $N^i$. $I^{i,j}$ is a current between nodes $N^i$ and $N^j$. At each node, according to Kirchoff's law,

$$\sum_{j=1}^{n_{neighbours}} I^{i,j} = 0 \tag{7}$$

where, $n_{neighbours}$ are the number of neighbour nodes to node $i$. According to Ohm's law,

$$I^{i,j} = \frac{\phi^i - \phi^j}{R^{i,j}} \tag{8}$$

where, $\phi^i$ is the electric potential at node $i$ and $R^{i,j}$ is the resistance between nodes $N^i$ and $N^j$.

Substituting Eq. 8 in Eq. 7,

At each interior node $i$,

$$\sum_{j=1}^{n_{neighbours}} \frac{\phi^i - \phi^j}{R^{i,j}} = 0 \tag{9}$$

At the boundary node $i$,

$$\sum_{j=1}^{n_{neighbours}} \frac{\phi^i - \phi^j}{R^{i,j}} = I^{total} \tag{10}$$

where, $I^{total}$ is the current flowing through the boundary (top or bottom) edge. Note that $\phi^i$ at the boundary is defined as a boundary condition. At each node, the Eqs. 9 or 10 are assembled to form a global system of linear equations of the form,

$$G\,\Phi = I \tag{11}$$

where, $G$ is the global conductivity matrix. $\Phi$ is the electric potential vector and $I$ is the current vector. Eq. 11 is solved to calculate $I^{total}$. The total resistance of the composite specimen, $R^{total}$ is calculated as, $R^{total} = (\phi^{top\ edge} - \phi^{bottom\ edge})/I^{total}$.

In this model, the best available analytical expression to calculate the resistance $R^{i,j}$ of each node pair is the one described in section 2.1 in Eq. 6. The total resistance of the composite specimen shown in Fig. 6a calculated using the Resistor Network method is 9428.5 Ω which shown a relative error of 28.3 %. It can be seen that approximation of the intra particle resistance leads to significant errors in the prediction of the resistance of the composite specimen as well.

## 3.3 Conditional Generative Adversarial Network

Generative Adversarial Networks are typically used to generate synthetic data that matches the dataset it is trained on [18]. It learns a mapping from a random noise $z$ to output image $y$, $G: z \to y$. A conditional Generative Adversarial Network (cGAN) learns a mapping from an input image $x$ and a random noise $z$ to $y$, $G: \{x, z\} \to y$ [19]. A cGAN is made up of two components – a generator and a discriminator. Generator G have a U-net architecture and discriminator D have a series of convolutional layers. Fig. 8 shows the overall architecture of the conditional GAN model.

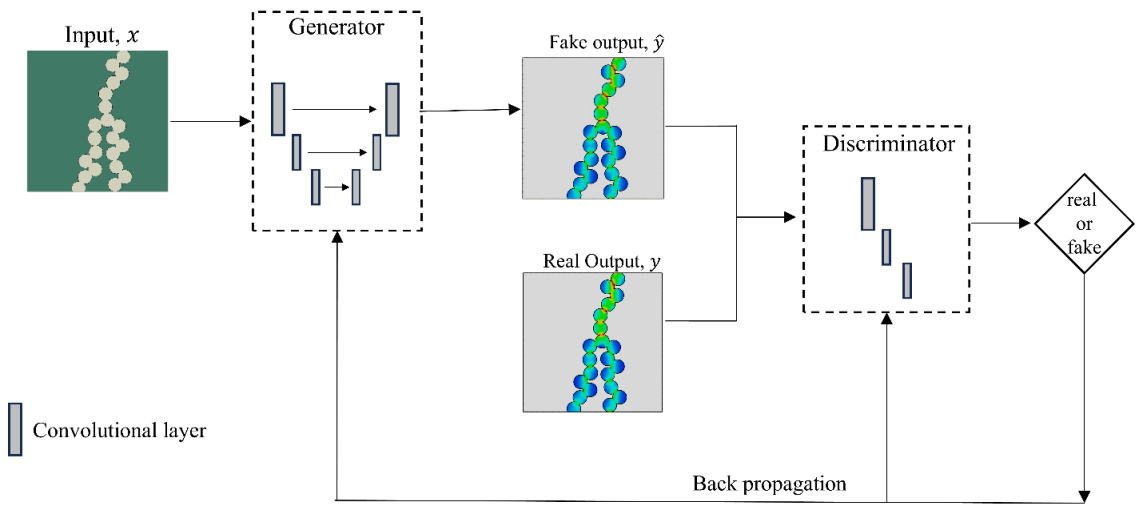

Fig.8. The process flow of a conditional GAN model implemented to determine current density in the composite specimen.

During the training process, an input image $x$ is fed to the generator that produces a fake output $\hat{y}$. The discriminator takes this fake output $\hat{y}$ and compares it with real output $y$ to check how close the fake output is to the real one. The Loss function of the entire model is,

$$\mathcal{L} = \mathcal{L}_{cGAN}(G,D) + \lambda\, \mathcal{L}_{L1}(G) \tag{12}$$

such that,

$$\mathcal{L}_{cGAN}(G,D) = \mathbb{E}_{x,y}[\log D(x,y)] + \mathbb{E}_{x,z}[\log(1 - D(G(x,y)))] \tag{13}$$

$$\mathcal{L}_{L1}(G) = \mathbb{E}_{x,y,z}[\|y - G(x,z)\|_1] \tag{14}$$

In this work, cGAN model is used to predict the current density of the 2D composite specimen studied here. The cGAN model was trained on 2000 images of composite specimens. Fig. 9 shows the prediction of the current density for 5 composite specimens (a-e) as calculated from FEM simulations (called as Ground Truth) and the prediction of cGAN (named as 'Prediction' in Fig.9) and a difference image that shows the difference in the current density values at each pixel position. It can be seen in the difference images that most of color contour in the particles in blue color. This indicates that the cGAN model was able to accurately replicate the current density contours as seen in the Ground Truth images.

Further, for the same specimens, Resistor Network models were created. Table. 1 shows the resistance of the 5 composite specimens (Fig.9a-e) as calculated from FEM results, the Resistor Network and the cGAN models. The resistance from the FEM and cGAN models was calculated by integrating current density at the same equipotential line for images of both methods (refer Fig.9). Then from the current value, the resistance was calculated from the potential difference. It can be seen that the % relative error for the cGAN model is much smaller than the Resistor Network model.

Table 2. Resistance of the composite samples as predicted by FEM, Resistor Network and cGAN models.

|  |  | Composite specimens |  |  |  |  |
| --- | --- | --- | --- | --- | --- | --- |
|  |  | a | b | c | d | e |
| FEM results | Resistance($\Omega$) | 18144.4 | 13236.7 | 14163.9 | 13316.3 | 13584.1 |
| Resistor Network model | Resistance($\Omega$) | 12864.5 | 9428.5 | 10023.0 | 9626.8 | 9761.6 |
|  | % relative error | 29.1 | 28.8 | 29.2 | 27.7 | 28.1 |
| cGAN model | Resistance($\Omega$) | 18968.0 | 13575.6 | 14586.7 | 13583.0 | 13887.5 |
|  | % relative error | 4.5 | 2.5 | 3.0 | 2 | 2.2 |

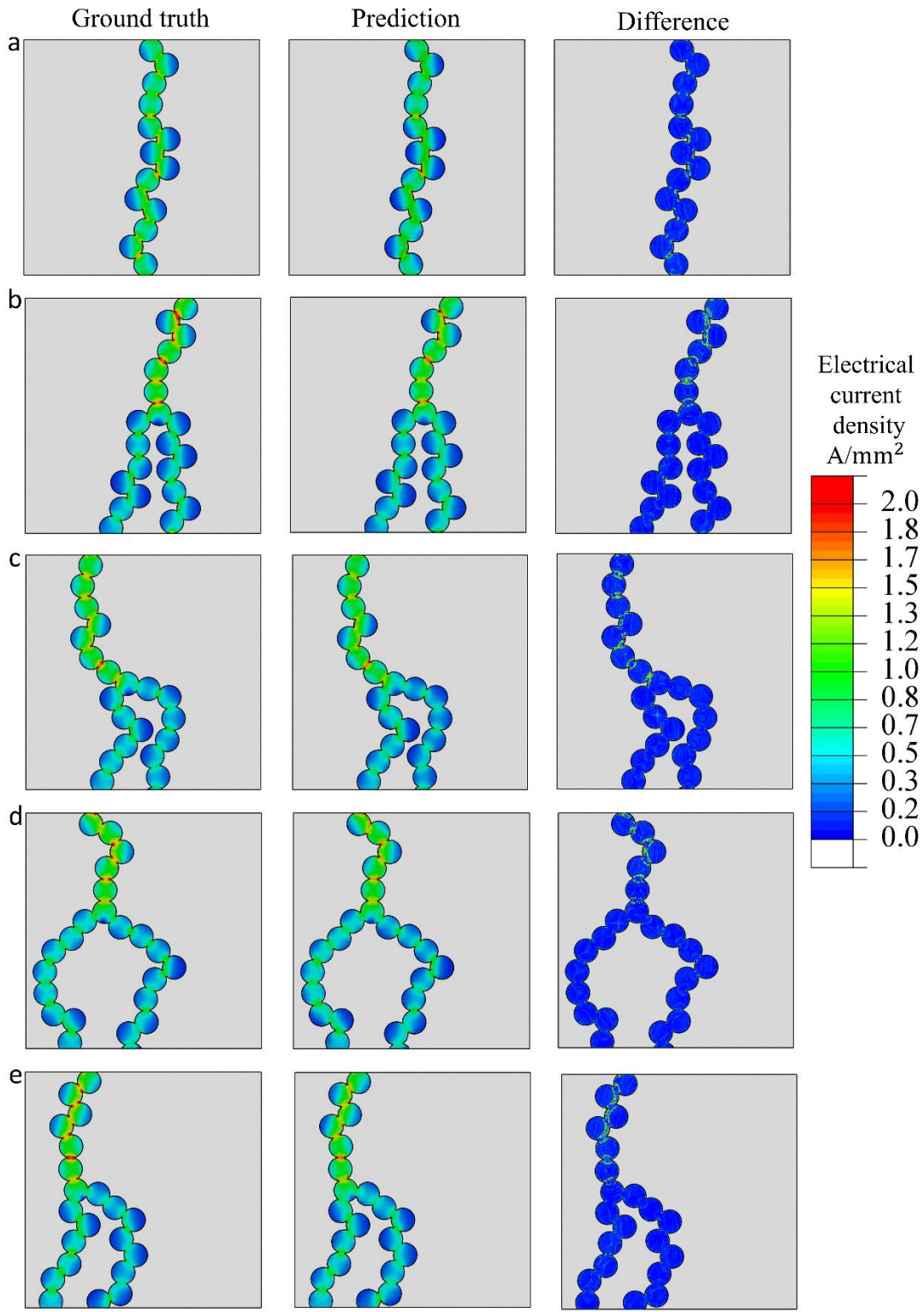

Fig.9. FEM (Ground Truth) and cGAN (Prediction) results of the prediction of current density in five (a-e) composite specimens.

## 4. Conclusion

The work focusses on modelling electrical resistance of 2D composite specimens made up of circular conductive particles embedded in polymer matrix. A typical method of modelling such systems is a Resistor Network method which is easily scalable but is inaccurate due to the way the particle resistance is modelled. The best available method of modelling particle resistance is analytical solutions to the Laplace equation. However, this method fails in the presence of complex inter-particle contact. This work demonstrates use of a generative AI method called cGAN to model the current flow in the composite specimens and shows that it gives much better accuracy then the Resistor Network method. In future studies, the cGAN method will be used to model larger microstructures to predict resistances at the scale of applications.